\def \PL {{ Phys. Lett.} }
\def \CMP {{ Commun. Math. Phys. }} 
\def \JMP {{ J. Math. Phys. }} 
\def \NP {{ Nucl. Phys.} }

\def \IJMP {{ Int. J. Mod. Phys.}}
\def \bc {\begin{center}}
\def \ec {\end{center}}

\def \bfr {\begin{flushright}}
\def \efr {\end{flushright}}

\def \ba{\begin{array}}
\def \ea{\end{array}}

\def \bea {\begin{eqnarray}}
\def \eea {\end{eqnarray}}

\def \be {\begin{equation}}
\def \ee {\end{equation}}

\def\nn{\nonumber}

\def\l[{\left[}
\def\r]{\right]}

\def\cL{{\cal L}}

\def\diff{{\rm diff}}
\def\sdiff{{\rm sdiff}}
\def\ba{\bar{a}}

\def\kbar{{\mathchar'26\mkern-9mu\lambda}}

\documentstyle[11pt,a4]{article}

\begin{document}

\centerline{}\vskip 1.5cm
\begin{center} 
{\large {\bf Structure Constants  
for New Infinite-Dimensional 
Lie Algebras of $U(N_+,N_-)$ Tensor Operators and Applications}}
\end{center}
\bigskip
\centerline{ {\sc M. Calixto}\footnote{E-mail: pymc@swansea.ac.uk 
/ calixto@ugr.es}  }
\bigskip

\begin{center}
{\it Department of Physics, University of Wales Swansea, Singleton Park, 
Swansea, SA2 8PP, U.K.}\\ and \\
{\it Instituto Carlos I de F\'\i sica Te\'orica y Computacional, Facultad
de Ciencias, Universidad de Granada, Campus de Fuentenueva, 
Granada 18002, Spain.} 
\end{center}

\bigskip

\bigskip
\begin{center}
{\bf Abstract}
\end{center}
\small          

\begin{list}{}{\setlength{\leftmargin}{3pc}\setlength{\rightmargin}{3pc}}
\item 
The structure constants for Moyal brackets of an infinite 
basis of functions on the 
algebraic manifolds $M$ of pseudo-unitary groups 
$U(N_+,N_-)$ are provided. They generalize the Virasoro and ${\cal W}_\infty$ 
algebras to higher dimensions. The connection with 
volume-preserving diffeomorphisms on $M$, higher generalized-spin  
and tensor operator algebras of $U(N_+,N_-)$ 
is discussed. These centrally-extended, infinite-dimensional 
Lie-algebras provide also the arena 
for non-linear integrable field theories in higher dimensions,  
residual gauge symmetries of higher-extended objects in the light-cone gauge 
and $C^*$-algebras for tractable non-commutative versions of symmetric 
curved spaces.
\end{list}

\normalsize


\newpage
The general study of infinite-dimensional algebras 
and groups, their quantum deformations (in particular, central extensions) 
and representation theory has not progressed very far, except for some 
important achievements in one- and two-dimensional systems, 
and there can be no doubt that 
a breakthrough in the subject would provide new insights into the two 
central problems of modern physics: unification of all interactions and exact 
solvability in QFT and statistics. 

The aforementioned achievements refer mainly to  
Virasoro and Kac-Moody symmetries (see e.g. \cite{Kac,Goddard}), 
which have played a fundamental role in 
the analysis and formulation of conformally-invariant 
(quantum and statistical) field theories in one and two dimensions, 
and systems in higher 
dimensions which in some essential respects are one- or two-dimensional 
(e.g. String Theory). Generalizations of the Virasoro symmetry, as the 
algebra $\diff(S^1)$ of reparametrisations of the circle, lead to 
the infinite-dimensional Lie algebras of 
area-preserving diffeomorphisms $\sdiff(\Sigma)$ of two-dimensional 
surfaces $\Sigma$. These algebras naturally appear as a 
residual gauge symmetry in the theory of relativistic membranes \cite{Hoppe}, 
which exhibits an intriguing connection with the quantum mechanics 
of space constant (e.g. vacuum configurations) $SU(N)$ Yang-Mills potentials  
in the limit $N\to\infty$ \cite{Floratos}; the 
argument that the internal symmetry space of the $U(\infty)$ pure 
Yang-Mills theory must be a functional space, actually the space of 
configurations of a string, was pointed out in Ref. \cite{Gervais}. 
Moreover, the ${\cal W}_\infty$ and ${\cal W}_{1+\infty}$ algebras 
of area-preserving 
diffeomorphisms of the cylinder \cite{cylinder} generalize the underlying 
Virasoro gauged symmetry of the light-cone two-dimensional induced 
gravity discovered by Polyakov \cite{Polyakov} by including all positive 
conformal-spin currents \cite{Pope}, and induced actions for 
these ${\cal W}$-gravity theories have been proposed 
\cite{wgravity,Nissimov}. Also, the 
${\cal W}_{1+\infty}$ (dynamical) symmetry has been identified by 
\cite{Cappelli} as the set of canonical 
transformations that leave invariant the 
Hamiltonian of a two-dimensional electron gas in a perpendicular 
magnetic field, and appears to be relevant in the classification of 
all the universality classes of {\it incompressible quantum fluids} and 
the identification of the quantum numbers of the excitations in the 
Quantum Hall Effect. Higher-spin symmetry algebras where introduced in 
\cite{Fradkin} and could provide a guiding principle towards the 
still unknown ``M-theory''.

It is remarkable that area-preserving diffeomorphisms, higher-spin and 
${\cal W}$ algebras can 
be seen as distinct members of a one-parameter family $\cL_\mu(su(2))$ 
---or the non-compact version $\cL_\mu(su(1,1))$--- of non-isomorphic 
\cite{Hoppe2} infinite-dimensional Lie-algebras 
of $SU(2)$ ---and $SU(1,1)$--- tensor 
operators, more precisely, the factor algebra 
$\cL_\mu(su(2))={\cal U}(su(2))/{\cal I}_\mu$ of the universal enveloping 
algebra ${\cal U}(su(2))$ by the ideal 
${\cal I}_\mu=(\hat{C}-\hbar^2\mu){\cal U}(su(2))$ 
generated by the Casimir operator $\hat{C}$ of $su(2)$ 
($\mu$ denotes an arbitrary complex number). 
The structure constants for $\cL_\mu(su(2))$ and 
$\cL_\mu(su(1,1))$ are well known for the Racah-Wigner basis of 
tensor operators \cite{Biedenharn}, and they can be written in terms of 
Clebsch-Gordan and (generalized) $6j$-symbols \cite{Hoppe,Pope,Fradkin2}. 
Another interesting feature of $\cL_\mu(su(2))$ is that, when $\mu$ 
coincides with the eigenvalue of $\hat{C}$ in an irrep $D_j$ of 
$SU(2)$, that is $\mu=j(j+1)$, there exists and ideal $\chi$ in 
$\cL_{\mu}(su(2))$ such that 
the quotient $\cL_{\mu}(su(2))/\chi\simeq sl(2j+1,C)$ or $su(2j+1)$, by 
taking a compact real form of the complex Lie algebra \cite{Burnside}. 
That is, for $\mu=j(j+1)$ the infinite-dimensional algebra $\cL_{\mu}(su(2))$ 
collapses to a finite-dimensional one. 
This fact was used in \cite{Hoppe} to approximate 
$\lim_{\stackrel{\mu\to\infty}{\hbar\to 0}}\cL_\mu(su(2))
\simeq \sdiff(S^2)$ 
by $su(N)|_{N\to\infty}$ (``large number of colors''). 

The generalization of these constructions to general unitary groups 
proves to be quite unwieldy, and a canonical classification of 
$U(N)$-tensor operators has, so far, been proven to exist only for 
$U(2)$ and $U(3)$ (see \cite{Biedenharn} and references therein). 
Tensor labeling is provided in these cases by the Gel'fand-Weyl pattern for 
vectors in the carrier space of the irreps of $U(N)$. 

In this letter, a quite appropriate basis of operators for 
$\cL_{\vec{\mu}}(u(N_+,N_-)),    
\vec{\mu}=(\mu_1,\dots,\mu_{N})$, $N \equiv N_++N_-$,  
is provided and the structure 
constants, for the particular case of the boson realization of 
quantum associative operatorial algebras on algebraic manifolds 
$M_{{}_{N_+N_-}}=U(N_+,N_-)/U(1)^{N}$, 
are calculated. The particular set of operators in  
${\cal U}(u(N_+,N_-))$ is the following:
\bea
\hat{L}^{I}_{|m|}&\equiv&
\prod_{\alpha}(\hat{G}_{\alpha\alpha})^{I_\alpha-(\sum_{\beta>\alpha}
|m_{\alpha\beta}|+\sum_{\beta<\alpha}|m_{\beta\alpha}|)/2}
\prod_{\alpha<\beta} (\hat{G}_{\alpha\beta})^{|m_{\alpha\beta}|} \nn\\
\hat{L}^{I}_{-|m|}&\equiv&
\prod_{\alpha}(\hat{G}_{\alpha\alpha})^{I_\alpha-(\sum_{\beta>\alpha}
|m_{\alpha\beta}|+\sum_{\beta<\alpha}|m_{\beta\alpha}|)/{2}}
\prod_{\alpha<\beta} (\hat{G}_{\beta\alpha})^{|m_{\alpha\beta}|}\label{auarop}
\eea  
where $\hat{G}_{\alpha\beta},\, \alpha,\beta=1,\dots,N$, 
are the $U(N_+,N_-)$ Lie-algebra (step) generators with commutation relations:
\be
\left[\hat{G}_{\alpha_1\beta_1},\hat{G}_{\alpha_2\beta_2}\right]=
\hbar(\eta_{\alpha_1\beta_2}\hat{G}_{\alpha_2\beta_1}-
\eta_{\alpha_2\beta_1}\hat{G}_{\alpha_1\beta_2})\,,\label{pun}
\ee 
and $\eta={\rm diag}(1,\stackrel{N_+}{\dots},1,-1,
\stackrel{N_-}{\dots},-1)$ is used to raise and lower indices;  
the upper (generalized spin) index $I\equiv(I_1,\dots,I_N)$ 
of $\hat{L}$ in (\ref{auarop}) represents 
a $N$-dimensional vector which, for the present, is taken to 
lie on an half-integral lattice; the lower index 
$m$ symbolizes a integral upper-triangular $N\times N$ matrix, and $|m|$ 
means absolute value of all its entries. Thus, the operators 
$\hat{L}^I_m$ are labeled by $N+N(N-1)/2=N(N+1)/2$ indices, 
in the same way as wave functions 
$\psi^I_m$ in the carrier space of irreps of $U(N)$. 
An implicit quotient by the ideal 
${\cal I}_{\vec{\mu}}=\prod_{j=1}^N(\hat{C}_j-\hbar^j\mu_j)
{\cal U}(u(N_+,N_-))$ generated by 
the Casimir operators 
\be
\hat{C}_1={\hat{G}_{\alpha}}^{\alpha}=\hbar\mu_1\,,\,\,\,\,
\hat{C}_2={\hat{G}_{\alpha}}^{\beta}
{\hat{G}_{\beta}}^{\alpha}=\hbar^2\mu_2\,,\,\dots\label{casimir}
\ee
is understood. The manifest expression of the structure constants $f$  
for the commutators 
\be
\l[\hat{L}^{I}_{m},\hat{L}^{J}_{n}\r]= 
\hat{L}^{I}_{m}\hat{L}^{J}_{n}-\hat{L}^{J}_{n}\hat{L}^{I}_{m}=
f^{IJl}_{mnK}[{\vec{\mu}}]\hat{L}^{K}_{l}\label{commu}
\ee
of a pair of operators (\ref{auarop}) of  $\cL_{\vec{\mu}}(u(N_+,N_-))$ 
entails an unpleasant and difficult computation, because of inherent 
ordering problems.  However, 
the essence of the full quantum algebra $\cL_{\vec{\mu}}(u(N_+,N_-))$ 
can be still captured in a classical construction by extending 
the Poisson-Lie bracket 
\be
\left\{L^{I}_{m},L^{J}_{n}\right\}_{{\rm PL}}=
(\eta_{\alpha_1\beta_2}
{G}_{\alpha_2\beta_1}-\eta_{\alpha_2\beta_1}{G}_{\alpha_1\beta_2})
\frac{\partial L^{I}_{m}}{\partial G_{\alpha_1\beta_1}}
\frac{\partial L^{J}_{n}}{\partial G_{\alpha_2\beta_2}}\label{poissonlie}
\ee
of a pair of functions $L^{I}_{m},L^{J}_{n}$ on the 
commuting coordinates $G_{\alpha\beta}$ 
to its deformed version, in the sense of Ref. \cite{Bayen}. 
To perform calculations with 
(\ref{poissonlie}) is still rather complicated because of non-canonical 
brackets for the generating elements $G_{\alpha\beta}$. Nevertheless, there 
is a standard boson operator realization 
$G_{\alpha\beta}\equiv a_\alpha\ba_\beta$ of the generators of $u(N_+,N_-)$ 
for which things simplify greatly. Indeed, we shall understand that the 
quotient by the ideal generated by polynomials $G_{\alpha_1\beta_1}
G_{\alpha_2\beta_2}-G_{\alpha_1\beta_2}G_{\alpha_2\beta_1}$ is taken, so that 
the Poisson-Lie bracket (\ref{poissonlie}) coincides with the standard 
Poisson bracket
\be
\left\{L^{I}_{m},L^{J}_{n}\right\}_{{\rm P}}=\eta_{\alpha\beta}\left(
\frac{\partial L^{I}_{m} }{\partial a_{\alpha}}
\frac{\partial L^{J}_{n}}{\partial \ba_{\beta}}-
\frac{\partial L^{I}_{m} }{\partial \ba_{\beta}}
\frac{\partial L^{J}_{n}}{\partial a_{\alpha}}\right)\label{poisson}
\ee
for the Heisenberg-Weyl algebra. There is basically only one possible 
deformation of the bracket (\ref{poisson}) ---corresponding to a 
normal ordering--- that fulfills the Jacobi 
identities \cite{Bayen}, which is the Moyal bracket \cite{Moyal}:
\be
\left\{L^{I}_{m},L^{J}_{n}\right\}_{{\rm M}}=
{L}^{I}_{m}*{L}^{J}_{n}-{L}^{J}_{n}*{L}^{I}_{m}=\sum_{r=0}^{\infty} 
2\frac{(\hbar/2)^{2r+1}}{(2r+1)!}P^{2r+1}(L^{I}_{m},L^{J}_{n})\,,\label{Moyal}
\ee
where $L*L'\equiv\exp(\frac{\hbar}{2} P)(L,L')$ is an invariant associative 
$*$-product and  
\be
P^r(L,L')\equiv\Upsilon_{\imath_1\jmath_1}\dots\Upsilon_{\imath_r\jmath_r}
\frac{\partial^r L}{\partial x_{\imath_1}\dots\partial 
x_{\imath_r}}\frac{\partial^rL'}{\partial x_{\jmath_1}\dots\partial 
x_{\jmath_r}}\,,\label{star}
\ee
with $x\equiv(a,\ba)$ and $\Upsilon_{2N\times 2N}\equiv
\left(\begin{array}{cc} 0 & \eta \\ 
-\eta &0\ea\right)$. We set $P^0(L,L')\equiv LL'$; 
see also that $P^1(L,L')=\{L,L'\}_{{\rm P}}$. Note 
the resemblance between the Moyal bracket (\ref{Moyal}) 
for {\it covariant symbols} 
$L^{I}_{m}$ and the standard commutator (\ref{commu}) for operators 
$\hat{L}^{I}_{m}$. It is worthwhile mentioning that Moyal brackets where 
identified as the primary quantum deformation ${\cal W}_\infty$ of the 
classical algebra $w_\infty$ of area-preserving diffeomorphysms of the 
cylinder (see Ref. \cite{Fairlie}).  

With this information at hand, the manifest expression 
of the structure constants $f$ for the Moyal bracket (\ref{Moyal}) is the 
following:
\bea
\left\{L^{I}_{m},L^{J}_{n}\right\}_{{\rm M}}&=& \sum_{r=0}^{\infty} 
2\frac{(\hbar/2)^{2r+1}}{(2r+1)!}\eta^{\alpha_0\alpha_0}\dots
\eta^{\alpha_{2r}\alpha_{2r}}f^{IJ}_{mn}(\alpha_0,\dots,\alpha_{2r})
L^{I+J-\sum_{j=0}^{2r} \delta_{\alpha_j}}_{m+n}\,,\nn\\
f^{IJ}_{mn}(\alpha_0,\dots,\alpha_{2r})&=& \sum_{\wp\in\Pi^{(2r+1)}_2}
(-1)^{\ell_\wp+1}\prod_{s=0}^{2r} f_\wp(I^{(s)}_{\alpha_{\wp(s)}},m)
f_\wp(J^{(s)}_{\alpha_{\wp(s)}},-n)\,,\label{chorizo}\\
f_\wp(I^{(s)}_{\alpha_{\wp(s)}},m)&=&I^{(s)}_{\alpha_{\wp(s)}}+
(-1)^{\theta(s-\ell_\wp)}(\sum_{\beta>\alpha_{\wp(s)}}m_{\alpha_{\wp(s)}\beta}-
\sum_{\beta<\alpha_{\wp(s)}}m_{\beta\alpha_{\wp(s)}})/2\,,\nn\\
I^{(s)}_{\alpha_{\wp(s)}}&=& I_{\alpha_{\wp(s)}}-\sum_{t=(\ell_\wp+1)
\theta(s-\ell_\wp)}^{s-1} \delta_{\alpha_{\wp(t)},\alpha_{\wp(s)}}\,,\;\;\;
I^{(0)}=I^{(\ell_\wp+1)}\equiv I\,,\nn\\
\theta(s-\ell_\wp)&=&\left\{\begin{array}{l} 0\,,\;\;s\leq\ell_\wp 
\\ 1\,,\;\;s>\ell_\wp\ea\right.\,,\;\;\;\;\delta_{\alpha_j}=
(\delta_{1,\alpha_j},\dots,\delta_{N,\alpha_j})\,,\nn
\eea
where $\Pi^{(2r+1)}_2$ denotes the set of all possible partitions $\wp$ of a 
string $(\alpha_0,\dots,\alpha_{2r})$ of length $2r+1$ into two substrings 
\be
(\overbrace{\alpha_{\wp(0)},\dots,\alpha_{\wp(\ell)}}^{\ell_\wp})
(\overbrace{\alpha_{\wp(\ell+1)},\dots,\alpha_{\wp(2r)}}^{2r+1-\ell_\wp})
\ee
of length $\ell_\wp$ and $2r+1-\ell_\wp$, respectively. The number of elements 
$\wp$ in $\Pi^{(2r+1)}_2$ is clearly ${\rm dim}(\Pi^{(2r+1)}_2)=
\sum_{\ell=0}^{2r+1}\frac{(2r+1)!}{(2r+1-\ell)!\ell !}=2^{2r+1}$. 

For $r=0$, there are just 2 partitions: $(\alpha)(\cdot),\, 
(\cdot)(\alpha)$, and the leading (classical, $\hbar\to 0$) structure 
constants are, for example:
\be
f^{IJ}_{mn}(\alpha)=J_{\alpha}
(\sum_{\beta>\alpha}m_{\alpha\beta}-
\sum_{\beta<\alpha}m_{\beta\alpha})
-I_{\alpha}(\sum_{\beta>\alpha}n_{\alpha\beta}-\sum_{\beta<\alpha}
n_{\beta\alpha})\,.\label{leading}
\ee
They reproduce in this limit 
the Virasoro commutation relations for the particular 
generators $V_k^{(\alpha\beta)}\equiv 
L^{\delta_\alpha}_{k e_{\alpha\beta}}$, where $k\in {\cal Z}$ and 
$e_{\alpha\beta}$ 
denotes an upper-triangular matrix with zero entries except for 
1 at $(\alpha\beta)$-position, if $\alpha 
< \beta$, or 1 at the $(\beta\alpha)$-position if $\alpha 
> \beta$. Indeed, there are $N(N-1)/2$ 
{\it non-commuting} Virasoro sectors in (\ref{chorizo}), corresponding to 
each positive root in $SU(N_+,N_-)$, 
with classical commutation relations:
\be
\left\{V_k^{(\alpha\beta)}, V_l^{(\alpha\beta)}\right\}_{\rm P}=
\eta^{\alpha\alpha}{\rm sign}(\beta-\alpha)\, (k-l)V_{k+l}^{(\alpha\beta)}\,.
\ee

For large $r$, we can benefit from the 
use of algebraic-computing programs like 
\cite{Mathematica} to deal with the high number of partitions. 

Our boson operator realization $G_{\alpha\beta}\equiv a_\alpha\ba_\beta$ 
of the $u(N_+,N_-)$ generators corresponds to the particular 
case of $\vec{\mu}_0=(N,0,\dots,0)$ for the Casimir eigenvalues, 
so that the commutation relations (\ref{chorizo}) are related to 
the particular algebra $\cL_{\vec{\mu}_0}(u(N_+,N_-))$ (see below for more 
general cases). A different (minimal) realization of $\cL_\mu(su(1,1))$ 
in terms of a single boson $(a,\bar{a})$, which corresponds to 
$\mu_c=\bar{s}_c(\bar{s}_c-1)=-3/16$ for the critical value $\bar{s}_c=3/4$ 
of the {\it symplin} degree of freedom $\bar{s}$, 
was given in \cite{symplin}; this case is also related to the {\it 
symplecton} algebra of \cite{Biedenharn}. Note the close resemblance 
between the algebra (\ref{chorizo}) 
---and the leading structure constants (\ref{leading})--- and the quantum 
deformation ${\cal W}_\infty\simeq \cL_0(su(1,1))$ of the algebra 
of area-preserving diffeomorphisms of the cylinder \cite{Pope,Fairlie}, 
although we recognize that the case discussed in this letter is far richer.

If the analyticity of the symbols $L^I_m$ of (\ref{auarop}) is 
taken into account, then one should worry about a restriction 
of the range of the indices $I_\alpha,m_{\alpha\beta}$. The subalgebra 
$\cL^\Lambda_{\vec{\mu}_0}(u(N_+,N_-))\equiv\{ L^I_m\,|\,\, 
\Lambda_\alpha=I_\alpha-(\sum_{\beta>\alpha}
|m_{\alpha\beta}|+\sum_{\beta<\alpha}|m_{\beta\alpha}|)/2\in 
{\cal N}\}$\footnote{${\cal N}$, ${\cal Z}$, $\Re$ and ${\cal C}$ 
denote the set of natural, integer, real and complex numbers, respectively} of 
polynomial functions on $G_{\alpha\beta}$, 
the structure constants $f^{IJ}_{mn}(\alpha_0,\dots,\alpha_{2r})$ of which 
are zero for  $r>(\sum_\alpha(I_\alpha+J_\alpha)-1)/2$, can be extended beyond 
the ``wedge'' $\Lambda\geq 0$ 
by analytic continuation, that is, by revoking this restriction to 
$\Lambda\in {\cal Z}/2$. 
The aforementioned ``extension beyond the wedge'' (see \cite{Pope,Fradkin} for 
similar concepts)  makes possible the existence of conjugated pairs  
$(L^I_m,L^{I'}_{-m})$, with 
$\sum_\alpha I_\alpha+I'_\alpha=2r+1$ and $I_\alpha+I'_\alpha\equiv 
r_\alpha\in {\cal N}$,  that give rise to central terms 
under commutation:
\be
\Xi({L}^I_m,{L}^{I'}_n)=\frac{\hbar^{2r+1}2^{-2r}(-1)^{\sum_{\alpha=N_++1}^N 
r_\alpha}}{\prod_{\alpha=1}^N
(2r+1-r_\alpha)!}
f^{II'}_{m,-m}(1^{(r_1)},\dots,N^{(r_N)})
 \delta_{m+n,0}\hat{1}\,,\label{twococycle}
\ee
where  $(1^{(r_1)},\dots,N^{(r_N)})$ is a string of length 
$\sum_{\alpha=1}^N r_\alpha=2r+1$ and $\alpha^{(r_\alpha)}$ 
denotes  a substring made of $r_\alpha$-times $\alpha$, for 
each $\alpha$. The generator $\hat{1}\equiv L^0_0$ 
is central (commutes 
with everything) and the Lie algebra two-cocycle (\ref{twococycle}) 
defines a non-trivial central extension of 
$\cL_{\vec{\mu}_0}(u(N_+,N_-))$ by $U(1)$.

A thorough study of the Lie-algebra 
cohomology of $\cL_{\vec{\mu}}(u(N_+,N_-))$ and its irreps 
still remains to be acoomplished; 
it requires a separate attention and shall be  
left for future works. Two-cocycles like (\ref{twococycle}) provide 
the essential ingredient to construct invariant geometric action functionals 
on coadjoint orbits of $\cL_{\vec{\mu}}(u(N_+,N_-))$ 
---see e.g. \cite{Nissimov} 
for the derivation of the WZNW action of $D=2$ matter fields coupled to 
chiral ${\cal W}_\infty$ gravity background 
from ${\cal W}_\infty\simeq\cL_0(su(1,1)))$. 

In order to deduce the structure constants for 
general $\cL_{\vec{\mu}}(u(N_+,N_-))$ from  
$\cL_{\vec{\mu}_0}(u(N_+,N_-))$, a procedure 
similar to that of Ref. \cite{Fradkin}, for the particular case of 
${\cal U}(sl(2,\Re))$, can be applied. Special attention must be paid 
to the limit $\lim_{\stackrel{\vec{\mu}\to\infty}{\hbar\to 0}}
\cL_{\vec{\mu}}(u(N_+,N_-))\simeq {\cal P}_{\cal C}(M_{{}_{N_+N_-}})$, which 
coincides with  the Poisson algebra of complex (wave) functions 
$\psi^I_{|m|},\, \psi^I_{-|m|}\equiv\bar{\psi}^I_{|m|}$ on algebraic 
manifolds (coadjoint orbits \cite{Kirillov}) 
$M_{{}_{N_+N_-}}\simeq U(N_+,N_-)/U(1)^{N}$.
\footnote{For $N_-\not=0$, other cases 
could  be also contemplated (e.g. continuous series of $SU(1,1)$)}  
It is well known that there exists 
a natural symplectic structure $(M_{{}_{N_+N_-}},\Omega)$,  
which defines the Poisson algebra
\be
\left\{\psi^I_m,\psi^J_n\right\}=\Omega^{\alpha_1\beta_1;\alpha_2\beta_2}
\frac{\partial\psi^I_m}{\partial g^{\alpha_1\beta_1}}
\frac{\partial\psi^J_n}{\partial g^{\alpha_2\beta_2}}\label{poiscoad}
\ee
and an invariant symmetric bilinear form 
$\langle\psi^I_m|\psi^J_n\rangle=
\int v(g)\bar{\psi}^I_m(g)\psi^J_n(g)$ 
given by the natural invariant measure $v(g)\sim \Omega^{N(N-1)/2}$ 
on $U(N_+,N_-)$, where $g^{\alpha\beta}= \bar{g}^{\beta\alpha}
 \in {\cal C},\,\alpha\not=\beta$, is a (local) 
system of complex coordinates on $M_{{}_{N_+N_-}}$.  
The structure constants for 
(\ref{poiscoad}) can be obtained through $f^{IJl}_{mnK}=
\langle\psi^K_l|\{\psi^I_m,\psi^J_n\}\rangle$. Also, an associative 
$\star$-product can be defined through the convolution of two functions 
$(\psi^I_m\star\psi^J_n)(g')\equiv\int v(g)\psi^I_m(g)
\psi^J_n(g^{-1}\bullet g')$, 
which gives the algebra ${\cal P}_{\cal C}(M_{{}_{N_+N_-}})$ a non-commutative 
character --- $g\bullet g'$ denotes the group composition 
law of $U(N_+,N_-)$. 
A manifest expression for  
all these structures is still in progress \cite{progress}. 

Taking advantage of all these geometrical tools, action functionals for 
$\cL_{\infty}(u(N_+,N_-))$ Yang-Mills gauge theories in $D$ dimensions 
could be built as: 
\bea
S&=&\int {\rm d}^Dx \langle F_{\nu\gamma}(x,g)|
F^{\nu\gamma}(x,g)\rangle\,,\nn\\
F_{\nu\gamma}&=&\partial_\nu A_\gamma-\partial_\gamma A_\nu +
\left\{A_\nu,A_\gamma\right\}\,,\\
A_\nu(x,g)&=&A_{\nu I}^m(x)\psi^I_m(g)\,,\;\;\nu,\gamma=1,\dots,D\,,\nn
\eea
the `vacuum configurations' (spacetime-constant potentials $X_\nu(g)\equiv 
A_\nu(0,g)$) of which, define the action for 
higher-extended objects: $N(N-1)$-`branes', 
in the usual nomenclature. Here, $\cL_{\infty}(u(N_+,N_-))$ plays the role 
of gauge symplectic (volume-preserving) diffeomorphisms 
$L_\psi\equiv\{\psi,\cdot\}$ on the $N(N-1)$-brane $M_{{}_{N_+N_-}}$. 
A particularly 
interesting case might be $SU(2,2)=U(2,2)/U(1)$: the conformal group in 
$3+1$ (or the AdS group in $4+1$) dimensions, in an attempt to construct 
`conformal gravities' in realistic dimensions. The infinite-dimensional 
algebra $\cL_\mu(u(2,2))$ 
might be seen as the {\it generalization of the Virasoro (two-dimensional) 
conformal symmetry to $3+1$ dimensions}.

Finally, let me comment on the potential relevance of the $C^*$-algebras
$\cL_{\vec{\mu}}({\cal G})$ on tractable non-commutative versions
\cite{Connes} of symmetric curved spaces $M=G/H$, where the notion of a
pure state $\psi^I_m$ replaces that of a point. The possibility of
describing phase-space physics in terms of the quantum analog of the
algebra of functions (the covariant symbols $L^I_m$), and the absence of
localization expressed by the Heisenberg uncertainty relation, was noticed
a long time ago by Dirac \cite{Dirac}. Just as the standard differential
geometry of $M$ can be described by using the algebra $C^\infty(M)$ of
smooth complex functions $\psi$ on $M$ (that is,
$\lim_{\stackrel{\vec{\mu}\to\infty}{\hbar\to 0}} \cL_{\vec{\mu}}({\cal
G})$, when considered as an associative, commutative algebra), a
non-commutative geometry for $M$ can be described by using the algebra
$\cL_{\vec{\mu}}({\cal G})$, seen as an associative algebra with a
non-commutative $*$-product like (\ref{Moyal},\ref{star}). The appealing
feature of a non-commutative space $M$ is that a $G$-invariant `lattice
structure' can be constructed in a natural way, a desirable property as
regards finite models of quantum gravity (see e.g. \cite{Madore} and Refs.
therein).  Indeed, as already mentioned, $\cL_{\vec{\mu}}({\cal G})$
collapses to ${\rm Mat}_{d}({\cal C})$ (the full matrix algebra of
$d\times d$ complex matrices) whenever $\mu_\alpha$ coincides with the
eigenvalue of $\hat{C}_\alpha$ in a $d$-dimensional irrep $D_{\vec{\mu}}$
of $G$. This fact provides a finite ($d$-points) `fuzzy' or `cellular'
description of the non-commutative space $M$, the classical (commutative)
case being recovered in the limit $\vec{\mu}\to\infty$. The notion of
space itself could be the collection of all of them, enclosed in a single
irrep of $\cL_{\vec{\mu}}({\cal G})$ for general $\vec{\mu}$, with
different multiplicities, as it actually happens with the reduction of an
irrep of the centrally-extended Virasoro group under its $SL(2,\Re)$
subgroup \cite{Jarama};  The multiplicity should increase with $\vec{\mu}$
(`the density of points'), so that classical-like spaces are more
abundant.  It is also a very important feature of
$\cL_{\vec{\mu}}(u(N_+,N_-))$ that the quantization deformation scheme
(\ref{Moyal}) does not affect the maximal finite-dimensional subalgebra
$su(N_+,N_-)$ (`good observables' or preferred coordinates \cite{Bayen})
of non-commuting `position operators' 
\bea
&y_{\alpha\beta}=\frac{\kbar}{2\hbar}(\hat{G}_{\alpha\beta}
+\hat{G}_{\beta\alpha})\,,\;\; 
y_{\beta\alpha}=\frac{i\kbar}{2\hbar}(\hat{G}_{\alpha\beta}
-\hat{G}_{\beta\alpha})\,,\;\;\;\;\alpha<\beta\,,&\nn\\
&y_\alpha=\frac{\kbar}{\hbar}(\eta_{\alpha\alpha}\hat{G}_{\alpha\alpha}-
\eta_{\alpha+1,\alpha+1}\hat{G}_{\alpha+1,\alpha+1})\,,& 
\eea 
on the algebraic manifold $M_{{}_{N_+N_-}}$, where $\kbar$ denotes a
parameter that gives $y$ dimensions of length (e.g., the square root of
the Planck area $\hbar {G}$).  The `volume' $v_j$ of the $N-1$
submanifolds $M_j$ of the {\it flag manifold} $M_{{}_{N_+N_-}}=
M_{N}\supset\dots\supset M_2$ (see e.g. \cite{Fulton} for a definition of
flag manifolds) is proportional to the eigenvalue $\mu_j$ of the
$su(N_+,N_-)$ Casimir operator $\hat{C}_{j}$ in those coordinates: 
$v_j=\kbar^{j}\mu_j$. Large volumes (flat-like spaces) correspond to a
high density of points (large $\mu$). In the classical limit $\kbar\to 0$,
$\mu\to \infty$, the coordinates $y$ commute. 

\section*{Acknowledgment}

I thank the University of Granada for a Post-doctoral grant and the 
Department of Physics of Swansea for its hospitality.


\begin{thebibliography}{99}
\bibitem{Kac} V.G. Kac, Infinite dimensional Lie algebras, Cambridge U.P., 
Cambridge (1985); \\
V.G. Kac, Bombay lectures on highest weight representations 
on infinite dimensional Lie algebras, World Scientific, Singapore (1987).
\bibitem{Goddard} P. Goddard and D.I. Olive, \IJMP {\bf A1}, 303 (1986).
\bibitem{Hoppe} J. Hoppe, MIT Ph.D. Thesis (1982); 
J. Hoppe, \IJMP {\bf A4}, 5235 (1989).
\bibitem{Floratos} E.G. Floratos, J. Iliopoulos and G. Tiktopoulos, 
\PL {\bf B217}, 285 (1989).
\bibitem{Gervais} J.L. Gervais and A. Neveu, \NP {\bf B192}, 463 (1981).
\bibitem{cylinder} I. Bakas, \PL {\bf B228}, 57 (1989).
\bibitem{Polyakov} A.M. Polyakov, Phys. Lett. {\bf B103}, 207 (1981). 
\bibitem{Pope} C.N. Pope, X. Shen and  L.J. Romans, \NP {\bf B339}, 191 (1990).
\bibitem{wgravity}E. Bergshoeff, C.N. Pope, L.J. Romans, 
E. Sezgin, X. Shen and 
K.S. Stelle, \PL {\bf B243}, 350 (1990).
\bibitem{Nissimov} E. Nissimov, S. Pacheva and I. Vaysburd, 
\PL {\bf B288}, 254 (1992). 
\bibitem{Cappelli} A. Cappelli and G.R. Zemba, 
\NP {\bf B490}, 595 (1997).
\bibitem{Fradkin} E.S. Fradkin and M.A. Vasiliev, 
Ann. Phys. (NY) {\bf 77}, 63 (1987).
\bibitem{Hoppe2} M. Bordemann, J. Hoppe and P. Schaller, \PL {\bf B232}, 199 
(1989).
\bibitem{Biedenharn} L.C. Biedenharn and J.D. Louck, Angular momentum 
in quantum physics, Addison-Wesley, Reading, MA (1981).\\
L.C. Biedenharn and J.D. Louck, The Racah-Wigner algebra in quantum theory, 
Addison-Wesley, New York, MA (1981).\\
L.C. Biedenharn and M.A. Lohe, Quantum group symmetry and q-tensor 
algebras, World Scientific, Singapore (1995).
\bibitem{Fradkin2} E.S. Fradkin and V.Y. Linetsky, \JMP {\bf 32}, 1218 (1991).
\bibitem{Burnside} N. Jacobson, Lectures in abstract algebra II, Van Nostrand, 
Princeton (1953).
\bibitem{Bayen} F. Bayen, M. Flato, C. Fronsdal, A. Lichnerowicz and 
D. Sternheimer, Ann. Phys.(NY) {\bf 111}, 61 (1978).
\bibitem{Moyal} J.E. Moyal, Proc. Cambridge Philos. Soc. {\bf 45}, 99 (1949).
\bibitem{Fairlie} D.B. Fairlie and J. Nuyts, \CMP {\bf 134}, 413 (1990).
\bibitem{Mathematica} S. Wolfram, Mathematica, Addison-Wesley, Reading, 
MA (1988).
\bibitem{symplin} M. Calixto, V. Aldaya and J. Guerrero, Int. J. Mod. 
Phys. {\bf A13}, {4889} (1998).
\bibitem{Kirillov} A. A. Kirillov, {\it Elements of the theory of 
representations}, Springer, Berlin (1976).
\bibitem{progress} M. Calixto, S. Howes and J.L. Jaramillo, in progress.
\bibitem{Connes} A. Connes, Noncommutative Geometry, Academic Press (1994).
\bibitem{Dirac} P.A.M. Dirac, Proc. Roy. Soc. {\bf A109}, 642 (1926);
 Proc. Camb. Phil. Soc. {\bf 23}, 412 (1926).
\bibitem{Madore} J. Madore, {\it Gravity on Fuzzy Space-time}, gr-qc/9709002.
\bibitem{Jarama} J. L. Jaramillo and V. Aldaya,  J. Phys. 
{\bf A32} (Math. Gen.), L503 (1999).
\bibitem{Fulton} W. Fulton and J. Harris, Representation Theory, 
Springer-Verlag, New York (1991).
\end{thebibliography}
\end{document}